\documentclass[preprint,prd,showpacs,showkeys,preprintnumbers,amsmath,amssymb]{revtex4}

\newcommand{\beq}{\begin{equation}}
\newcommand{\eeq}{\end{equation}}
\newcommand{\lab}{\label}

\newcommand{\bfxi}{\mbox{\boldmath $\xi$}}
\newcommand{\bfom}{\mbox{\boldmath $\omega$}}

\newcommand{\bfx}{ {\bf x} }

\begin{document}

\preprint{}

\title{Gravitational Faraday rotation in a weak gravitational field}

\author{Mauro Sereno}
\altaffiliation[Also at ]{Osservatorio Astronomico di Capodimonte -
INAF, Salita Moiariello, 16, 80131 Napoli, Italia}
\email{sereno@na.infn.it}
\affiliation{Dipartimento di Scienze Fisiche, Universit\`{a}
degli Studi di Napoli ``Federico II", \\ and \\ Istituto Nazionale di
Fisica Nucleare, Sez. Napoli, \\ via Cinthia, Compl. Univ. Monte S.
Angelo, 80126 Napoli, Italia }

\date{December 10, 2003}

\begin{abstract}
\noindent We examine the rotation of the plane of polarization for linearly
polarized light rays by the weak gravitational field of an isolated
physical system. Based on the rotation of inertial frames, we review
the general integral expression for the net rotation. We apply this
formula, analogue to the usual electromagnetic Faraday effect, to some
interesting astrophysical systems: uniformly shifting mass monopoles
and a spinning external shell.
\end{abstract}

\maketitle

\section{Introduction}

Electromagnetic theory in a curved space-time, in the approximation of
geometric optics, provides some of the most well-known and stringent
tests of Einstein's general theory of gravitation. Under geometric
optics, a ray follows a null geodesic regardless of its polarization
state and the polarization vector is parallel transported along the
ray \cite{mtw73}. In the last decades, observations of both bending of
light and gravitational time delay have revealed themselves as a
powerful tool in observational astrophysics and cosmology. These
phenomena are fully accounted for in the gravitational lensing theory
\cite{pet+al01,sef}. On the other hand, effects of polarization along
the light path have not yet been measured.

The polarization vector of a linearly polarized electromagnetic wave
rotates due to the properties of the space-time. The gravitational
rotation of the plane of polarization in stationary space-times is a
gravitational analogue of the electro-magnetic Faraday effect, i.e.,
the rotation that a light ray undergoes when passing through plasma in
the presence of a magnetic field. The analogy was first noted in
\cite{pi+sa85}, where the problem of nonlinear interaction on
gravitational radiation was considered.

The first discussion of this relativistic effect goes back to 1957,
when Skrotskii \cite{skr57} applied a method previously developed by
Rytov \cite{ryt38} to consider geometric optics in a curved
space-time. For this historical reason, the gravitational effect on
the polarization of light rays is also known as Skrotskii or Rytov
effect. In 1958, Balazs~\cite{bal58} further stressed how the
gravitational field of a rotating body may behave as an optically
active medium. In 1960, Plebanski~\cite{ple60} solved the Maxwell's
equations in the gravitational field of an isolated physical system.
He showed how the polarization vector changes its direction due to the
deflection of the light ray, and, in addition to this change, how a
rotation of the plane of polarization around the propagation vector
may occur. Ten years later, Godfrey~\cite{god70} took a very different
approach. Following Mach's principle, he considered dragging of
inertial frames along with a rotating body and obtained an approximate
expression for the rotation of the polarization vector of a light ray
propagating along the rotation axis of a Kerr black hole. Trajectories
initially propagating parallel to the symmetry axis of a central
spinning body were studied in \cite{su+ma80}, where the problem was
formulated in a cylindrical-like Kerr solution. A different situation
was considered in \cite{con+al80}, where it was discussed how the
polarization features of X-ray radiation emitted from an accretion
disk surrounding a rotating black hole are also strongly affected by
general-relativistic effects. The relativistic rotation of the plane
of polarization was further studied in \cite{ko+ma02}. Solving the
equations of motion of a light ray in the first post-Minkowskian
approximation, a formula describing the Skrotskii effect for arbitrary
translational and rotational motion of gravitating bodies was derived.

Finally, the Skrotskii effect on light rays propagating in the vacuum
region outside the event horizon of a Kerr black hole has been
discussed in \cite{ish+al88,nou99}. In particular, the formulation in
\cite{nou99} stressed in an illuminating way the analogy with the
usual Faraday effect.

In this paper, we explore the gravitational Faraday rotation by the
gravitational field of an isolated system (lens) when the source of
radiation and the observer are remote from the gravitational lens. We
restrict to the weak gravitational field far from the lens, and
analyze it using linearized theory. This approximation holds for
almost all gravitational lensing phenomena. We consider gravitational
Faraday rotation by usual astrophysical systems, such as a system of
shifting stars acting as lenses or a galaxy deflecting light rays
emitted from background sources.

The paper is as follows. In Section~\ref{deri}, we extend the argument
of Godfrey~\cite{god70} on dragging of inertial frames to reobtain the
well known general formula for the angle of rotation of the plane of
polarization of a linearly polarized electromagnetic wave in a
stationary space-time. This heuristic approach allows us to face the
problem without integrating the equation of motion. In
Sec.~\ref{weak}, the weak-field, slow motion approximation is
introduced and the weak field limit of the gravitational Faraday
rotation is performed. In Section~\ref{appl}, we evaluate the Faraday
rotation for some systems of astrophysical interest. We examine a
system of uniformly moving lenses and a rotating external shell.
Section~\ref{conc} is devoted to some final considerations.

\section{Derivation of the gravitational Faraday rotation}
\label{deri}

Let us consider a stationary space-time embedded with a metric
$g_{\alpha
\beta}$ \footnote{Latin indices run from 1 to 3, whereas Greek indices
run from 0 to 3.}. Such
a metric can be written as \cite{la+li85}
\begin{equation}
ds^2 = h \left( dx^0-A_i dx^i \right)^2- d l_{\rm P}^2
\end{equation}
where we have introduced the notation
\begin{equation}
\label{far2}
h \equiv g_{00}, \ A_i \equiv -\frac{g_{0i}}{g_{00}},
\end{equation}
and
$$
{d l_{\rm P}}^2 \equiv \left( -g_{ij}
+\frac{g_{0i}g_{0j}}{g_{00}}\right)dx^i d x^j \equiv \gamma_{i j} dx^i
d x^j
$$
is the spatial distance in terms of the spatial metric $\gamma_{ij}$
\cite{la+li85}. As suggested in \cite{god70}, the rotation of the
photon's plane of polarization may be computed using an argument on
the dragging of inertial frames. Due to the gravito-magnetic field,
inertial frames near moving bodies are dragged in the direction of
motion of the sources of the gravitational field. A differential
rotation between adjacent frames results from the variation of the
rate of dragging with position. The angular velocity with which a
local inertial frame, instantaneously at rest in a stationary frame,
rotates relative to the stationary frame can be expressed as
\cite{la+li85},
\begin{equation}
\label{far1}
\mbox{\boldmath $\Omega$}^{\rm Sk} = -\frac{1}{2} \sqrt{h} \nabla {\times} \vec{\bf A},
\end{equation}
Following \cite{god70}, we assume the polarization vector to be
dragged along by the rotation of the inertial frames. This view agrees
with the result in \cite{mas75}. The net angle of rotation around the
tangent three-vector $\hat{\bf k}$ \footnote{The controvariant
components of spatial three-vectors are equal to the spatial
components of the corresponding four-vectors. Operations on such
three-vectors are defined in the three-dimensional space with metric
$\gamma_{\alpha
\beta}$.} along the path between the source and
the observer turns out to be
\begin{equation}
\label{far3}
\Omega^{\rm Sk} = - \frac{1}{2} \int_{\rm sou}^{\rm obs} \sqrt{h} \nabla {\times} \vec{\bf A} {\cdot} {\bf d}\vec{\bf x},
\end{equation}
where ${\bf d}\vec{\bf x}=\hat{\bf k}dl_{\rm P} $, in agreement with
equation~(20) in \cite{nou99}. Identifying $\nabla {\times} \vec{\bf A}$ with
the gravito-magnetic field $\vec{\bf B}_{\rm g}$ \cite{nou99}, we see
how Eq.~(\ref{far3}) takes the same form of the usual Faraday effect,
i.e., it is proportional to the integral of the component of the
gravito-magnetic field along the propagation path. The analogy is only
formal since the origins of the two effects are completely different.
The gravitational rotation of the plane of polarization is a purely
geometrical effect due to the structure of space-time, and no
frequency dependence occurs \cite{ish+al88,nou99}.

\section{The weak field limit}
\label{weak}

Standard hypotheses of gravitational lensing \cite{pet+al01,sef}
assume that the gravitational lens is localized in a very small region
of the sky and its lensing effect is weak. The deflector changes its
position slowly with respect to the coordinate system, i.e., the
matter velocity is much less than $c$, the speed of light; matter
stresses are also small (the pressure is much smaller than the energy
density). In this weak field regime and slow motion approximation,
space-time is nearly flat near the lens. As can be seen from
Eq.~(\ref{far3}), the order of approximation is determined by
off-diagonal components of the metric. To our aim, it is enough to
write
\begin{equation}
\lab{wf1}
ds^2 \simeq \left( 1+2\frac{\phi}{c^2} + {\cal{O}}(\varepsilon^4)
\right)c^2dt^2-8c dt \frac{
\vec{\bf V} {\cdot} d \vec{\bfx}}{c^3}-
\left( 1-2\frac{\phi}{c^2} + {\cal{O}}(\varepsilon^4) \right) d\vec{\bfx}^2,
\end{equation}
where $\varepsilon \ll 1$ denotes the order of approximation. $\phi$ is
the Newtonian potential,
\begin{equation}
\lab{wf2}
\phi (t, \vec{\bfx}) \simeq -G \int_{\Re^3} \frac{\rho (t, \vec{\bfx}^{'})}{ | \vec{\bfx} -
\vec{\bfx}^{'}|} d^3 x^{'};
\end{equation}
$\phi/c^2$ is of order $\sim {\cal O}(\varepsilon^2)$.

$\vec{\bf V}$ is a vector potential taking into account the
gravito-magnetic field produced by mass currents. To
${\cal{O}}(\varepsilon^3)$,
\begin{equation}
\lab{wf3}
\vec{\bf V} (t, \vec{\bfx}) \simeq -G \int_{\Re^3} \frac{(\rho \vec{\bf v})(t,
\vec{\bfx}^{'})}{ | \vec{\bfx}
-\vec{\bfx}^{'}|} d^3 x^{'},
\end{equation}
where $\vec{\bf v}$ is the velocity field of the mass elements of the
deflector.

We assume that, during the time light rays take to traverse the lens,
the potentials in Eqs.~(\ref{wf2},~\ref{wf3}) vary negligibly little.
Then, the lens can be treated as stationary. In
Eqs.~(\ref{wf2},~\ref{wf3}), we have neglected the retardation
\cite{sef}.

In the weak field limit, $h$ and $\vec{\bf A}$ are simply related to
the gravitational potentials. It is
\begin{eqnarray}
h & \simeq & 1+2 \frac{\phi}{c^2} + {\cal{O}}(\varepsilon^4),
\label{far5} \\ A_i & \simeq &
\frac{4}{c^3}V_i  + {\cal{O}}(\varepsilon^5) \label{far6};
\end{eqnarray}
the proper arc length reads
\begin{equation}
\label{delppN1}
d l_{\rm P} \simeq \left\{ 1- \frac{\phi}{c^2} + {\cal O}
(\varepsilon^3) \right\} d l_{\rm E},
\end{equation}
where $d l_{\rm E} \equiv \sqrt{ \delta_{ij}d x^i d x^j}$ is the
Euclidean arc length.

Inserting Eqs.~(\ref{far5},~\ref{far6},~\ref{delppN1}) in
Eq.~(\ref{far3}), we get
\begin{equation}
\Omega^{\rm Sk}  \simeq  -\frac{2}{c^3} \int_{p}  \nabla {\times} \vec{\bf V} {\cdot} {\bf \hat{k}} \ d l_{\rm E}
 + {\cal O}(\varepsilon^5),
\end{equation}
where $p$ is the spatial projection of the null geodesics and
$\hat{\bf k}$ is the unit tangent vector.

It is useful to employ the spatial orthogonal coordinates $(l, \xi_1,
\xi_2) \equiv (l, \bfxi)$, centred on the lens and such that the $l$-axis is along the
incoming, unperturbed light ray direction $\vec{\bf e}_{\rm in}$. The
lens plane, $(\xi_1,\xi_2)$, corresponds to $l=0$. The
three-dimensional position vector to the light ray $\vec{\bfx}$ can be
written as $\vec{\bfx} = \mbox{\boldmath $\xi$} + l \vec{\bf e}_{\rm
in}$.

To calculate the Skrotskii effect to order ${\cal O}(\varepsilon^3)$
we can adopt the Born approximation, which assumes that rays of
electromagnetic radiation propagate along straight lines, i.e, the
bending of the ray may be neglected. The integration along the line of
sight (l.o.s.) is accurate enough to evaluate the main contribution to
the net rotation \cite{io03prd}. To this order, we can employ the
unperturbed Minkowski metric $\eta_{\alpha
\beta}=(1,-1,-1,-1)$ and a constant unit propagation vector of the signal, ${\bf
\hat{k}}_{(0)}= (1,0,0)$.

The Faraday rotation to order ${\cal O} (\varepsilon^3)$ reads
\begin{equation}
\Omega^{\rm Sk}  \simeq -\frac{2}{c^3} \int_{\rm l.o.s.}
\left. \nabla {\times} \vec{\bf V} \right|_{\rm l.o.s.} d l_{\rm E} + {\cal O} (\varepsilon^5).
\label{far8}
\end{equation}
Eq.~(\ref{far8}) gives the main contribution to the gravitational
Faraday rotation.

\section{Applications}
\label{appl}

In this section, we provide explicit formulas for the gravitational
Faraday rotation for some relevant astrophysical systems.

\subsection{System of shifting lenses}

A system of point-like lenses moving with constant velocities induces
a gravitational Faraday rotation \cite{ko+ma02,ple60}. Stars in
motion, where the velocities can be treated as slow, provide a
suitable representation of such a system.

The gravito-magnetic potential generated by a point-like lens of mass
$M_i$ with vector position $\vec{\bfx}_i$, shifting with a constant
velocity $\vec{\bf v}_i$, reduces to
\begin{equation}
\vec{\bf V}_i ( \vec{\bfx})\simeq -G M_i \frac{\vec{\bf v}_i}{|\vec{\bfx} -\vec{\bfx}_i|}.
\end{equation}

The rotation angle from a system of $N$ shifting lenses is
\begin{eqnarray}
\Omega^{\rm Sk} & \simeq &  -\frac{2}{c^3} \sum_i^N  \int_{\rm l.o.s.} \left. \nabla {\times}
\vec{\bf V}_i \right|_{\rm l.o.s.} d l_{\rm E} + {\cal O} (\varepsilon^5) \nonumber
\\
& = & -\frac{4 G}{c^3} \sum_i^N  M_i
 \frac{\Delta \xi_{(i)1} v_{(i)2}-\Delta \xi_{(i)2} v_{(i)1}}{| \Delta \bfxi_{(i)}|^2} + {\cal O} (\varepsilon^5) \nonumber \\
\end{eqnarray}
where $\Delta \bfxi_{(i)} \equiv \bfxi - \bfxi_{(i)}$. To the lowest
order, only the projection along the line-of-sight of the total
angular momentum enters the effect.

\subsection{Rotating shell}

The gravito-magnetic potential takes a very simple form in the case of
a spherically symmetric distribution of matter in rigid rotation. We
limit to a slow rotation so that the deformation caused by rotation is
negligible and the body has a nearly spherical symmetry. Taking the
centre of the source as the spatial origin of a background inertial
frame, we get
\begin{eqnarray}
\label{grp1}
\vec{\bf V}  & \simeq & - \frac{4 \pi}{3} G \left\{ \frac{1}{x^3}
\int_0^x \rho (r)r^4 dr + \int_x^{+\infty} \rho (r)r dr
\right\} \vec{\bfom} {\times} \vec{\bfx} \nonumber
\\
& = & -\frac{G}{2} \vec{\bf J}(x){\times} \frac{\vec{\bfx} }{x^3}- \frac{4
\pi}{3} G \left(
\int_x^{+\infty} \rho (r)r dr
\right) \vec{\bfom} {\times} \vec{\bfx},
\end{eqnarray}
where $\vec{\bfom} = const.$ is the angular velocity and $\vec{\bf
J}(x) = \frac{8 \pi}{3}
\left( \int_0^x
\rho (r)r^4 dr \right) \vec{\bfom}$ is the angular momentum contributed from the matter within a radius $x \equiv |\vec{\bf x}|$.

Einstein's gravitational theory predicts peculiar phenomena inside a
rotating shell. It is interesting to calculate the gravito-magnetic
potential for such a system. The gravito-magnetic potential in
Equation~(\ref{grp1}), inside a uniform spherical shell of mass $M$,
radius $R$ and rotating with constant frequency, reduces to (see also
\cite{ciu+al03})
\begin{equation}
\vec{\bf V}^{\rm In}( \vec{\bfx)} \simeq -\frac{G M}{3 R}
\vec{\bfom} {\times} \vec{\bfx}.
\end{equation}
Outside the rotating shell ($x >R$),
\begin{equation}
\vec{\bf V}^{\rm Out}( \vec{\bfx)} \simeq -\frac{G M R^2}{3}
\frac{\vec{\bfom} {\times} \vec{\bfx}}{x^3}
= -\frac{G}{2}\frac{ \vec{\bf J} {\times} \vec{\bfx} }{x^3},
\end{equation}
where $\vec{\bf J} = \frac{2}{3} M R^2 \vec{\bfom}$.

 It is
\begin{eqnarray}
\nabla {\times} \vec{\bf V}^{\rm In}( \vec{\bfx}) & \simeq & -\frac{2 G M}{3 R}
\vec{\bfom} \\
\nabla {\times} \vec{\bf V}^{\rm Out}( \vec{\bfx}) & \simeq &
\frac{G}{2}\left[ \frac{\vec{\bf J} -
3( \vec{\bf J} {\cdot}\hat{\bfx} )\hat{\bfx}}{x^3} \right]
\end{eqnarray}
Let us consider a light ray which enters the shell, i.e. with impact
parameter $\bfxi \leq R$; the light ray enters and leaves the sphere
at $l_{\rm In}=
-\sqrt{ R^2- \xi^2}$ and $l_{\rm Out} = +\sqrt{ R^2- \xi^2}$, respectively.
The net rotation of the polarization vector is
\begin{eqnarray}
\Omega^{\rm Sk} & \simeq & - \frac{2}{c^3} \left\{ \int_{-\infty}^{l_{\rm In}} \left. \nabla {\times}
\vec{\bf V}^{\rm Out} \right|_{\rm l.o.s.} d l_{\rm E} +
\int_{l_{\rm In}}^{l_{\rm Out}} \left. \nabla {\times}
\vec{\bf V}^{\rm In} \right|_{\rm l.o.s.} d l_{\rm E} +
\int_{l_{\rm Out}}^{+\infty} \left. \nabla {\times}
\vec{\bf V}^{\rm Out} \right|_{\rm l.o.s.} d l_{\rm E}\right\}
+{\cal O}(\varepsilon^5) \nonumber
\\
& = &  \frac{4 G M}{c^3} \omega_{\rm l.o.s.} \sqrt{1-\left(
\frac{\xi}{R} \right)^2} +{\cal O}(\varepsilon^5). \label{she1}
\end{eqnarray}
The result vanishes if the angular velocity lies in the lens plane. Since
the gravitational Faraday rotation outside a rotating body, when
the light path does not enter the lens, is $\sim \frac{G^2 M_{\rm
TOT}}{c^5}\frac{J_{\rm l.o.s.}}{\xi^3}$, i.e, of order ${\cal
O}(\varepsilon^5)$ \cite{ish+al88,nou99,io03far}, the effect on the
light ray can be neglected at this order of approximation.

The case of a rotating external sphere of finite thickness can be
easily solved just integrating the result in Eq.~(\ref{she1}). Every
infinitesimal shell of radius $r^{'}$ with mass $dM= 4 \pi \rho
(r^{'})r^{'2} d r^{'} $ and angular velocity $\omega (r^{'})$
contributes an angle
\begin{equation}
d \Omega^{\rm Sk} \simeq  \frac{16 \pi G}{c^3} \rho (r^{'})
\omega_{\rm l.o.s.} (r^{'})
\sqrt{r^{'2}- \xi^2} r^{'} d r^{'} +{\cal O}(\varepsilon^5).
\end{equation}
Integrating from the impact parameter, $\xi$, to the external shell
radius $R$, we get the total gravitational Faraday rotation which a
light ray undergoes because of the spin of the external shell. We get
\begin{eqnarray}
\Omega^{\rm Sk} & = & \int d \Omega^{\rm Sk} \nonumber \\
& \simeq & \frac{16 \pi G}{c^3} \int_{\xi}^{R} \rho (r^{'})
\omega_{\rm l.o.s.} (r^{'})
\sqrt{r^{'2}- \xi^2} r^{'} d r^{'} +{\cal O}(\varepsilon^5).
\end{eqnarray}

Let us consider a homogeneous sphere of constant density in rigid
rotation. The plane of polarization of a light ray, that penetrates
through this rotating body, is rotated of
\begin{eqnarray}
\Omega^{\rm Sk} & \simeq & \frac{16 \pi G}{3 c^3} \rho \ \omega_{\rm l.o.s.}
\left( R^2 -\xi^2 \right)^{3/2} +{\cal O}(\varepsilon^5) \nonumber
\\
& =& \frac{10 G}{c^3} J_{\rm l.o.s.} \frac{\left( R^2 -\xi^2
\right)^{3/2}}{R^5} +{\cal O}(\varepsilon^5), \label{hom1}
\end{eqnarray}
where $J_{\rm l.o.s.}$ is the component along the line of sight of the
total angular momentum of the sphere (see also \cite{ple60}).

\section{Conclusions}
\label{conc}
We have discussed the theory of the gravitational Faraday rotation in
the weak field limit. To the lowest order of approximation, only the
projection along the line-of-sight of the total angular momentum
contributes to the Skrotskii effect.

The rotation angle of the plane of polarization of a linearly
polarized electromagnetic wave is of order ${\cal O}(\varepsilon^3)$
in both cases of a system of shifting sources and inside a rotating
shell. These models suitably apply to well known gravitational lensing
systems. During microlensing events on the Galactic scale, a star,
acting as deflector, moves relatively to a background source. This is
the case of a shifting lens.

A distant quasar lensed by a foreground galaxy may form images inside
the galaxy radius. In such an astrophysical configuration, photons
propagate inside a rotating shell. Since the Faraday rotation due to
external rotating shell is of order ${\cal O}(\varepsilon^3)$, it
could induce a detectable effect. High quality data in total flux
density, percentage polarization and polarization position angle at
radio frequencies already exist for multiple images of some
gravitational lensing systems, like B0218+357 \cite{big+al99}.

The prospects to detect the gravitational Faraday rotation will be the
argument of a forthcoming paper.

\begin{acknowledgments}
The author wishes to thank the Dipartimento di Fisica ``E.
Caianiello", Universit\`{a} di Salerno, Italia, for the hospitality when
he first worked on the idea illustrated in the paper.
\end{acknowledgments}

\end{document}